# Ensemble methods for solving problems of medical diagnosis

*Anastasia Grigoreva* [1], *Andrey Trufanov* [1] *and Stanislav Grigorev* [1]

[1] *Irkutsk, Irkutsk National Research Technical University*

**ABSTRACT**

A consolidating method for analyzing series of observations based on a fitted model of a mixture of catalysts of the main components is proposed, which makes it possible to study any number of markers. Contrasting the longitudinal approach, it eliminates the need to connect regression analysis methods with their own uncertainties when choosing particular models. The consolidating method allows obtaining an original result in the subject area of early diagnosis of a disease: all options for using markers demonstrate an increase in classification accuracy with an increase in the length of a series of examinations.

**Introduction.** Ensemble methods are a fairly powerful tool for building a set of machine learning models for the same task to obtain the best indicators of adequacy, accuracy and efficiency of application for data series predictions. In addition, in the absence of "strong learners" to train the model, the use of ensemble methods may be the only correct solution for the task. In this work, studies were carried out on the construction and evaluation of the effectiveness of models by various classification methods for detecting the diagnosis of oncological diseases in the early stages [1].

**Data collection.** Biomarkers can be used to detect disease early, before it becomes clinically evident. Cancer biomarkers are used to detect cancer. Such markers can be obtained on the basis of the so-called tumor microchip - a biochemical analysis based on the principle of forming an array of markers on a microscope slide. Thus, in particular, for the detection of pancreatic cancer, the data of observation of a tumor marker based on the CA-19-9 marker microchip are used [2].

The possibility of presenting and processing the results of serial observations of various markers was studied in the interests of solving the problem of diagnosing oncological diseases. For the experiments, we used a set formed on the basis of data from the early stage of cancer diagnosis, obtained in one of the regional clinics in the Irkutsk region for 10 years.

The studied data set contained the results of observations in 71 cases of oncological diseases of various nature (conditional diagnosis D = 1) and for 70 healthy patients from the control group (D = 0) (Table 1).

Table 1. Frequency of occurrence of series of observations of length L

| D | L | | | | | | | | | | |
|---|---|---|---|---|---|---|---|---|---|---|---|
|   | **1** | **2** | **3** | **4** | **5** | **6** | **7** | **8** | **9** | **10** | **∑** |
| 0 | 0 | 2 | 6 | 9 | 4 | 12 | 10 | 9 | 18 | 0 | 70 |
| 1 | 14 | 7 | 24 | 9 | 10 | 6 | 1 | 0 | 0 | 0 | 71 |

The data were broken down into series of observations for individual patients. The length L of the series varied from one patient to another. Two types of markers were considered: common and free. There were 683 observations in total, each of which included patient ID, diagnosis, time to final diagnosis, biomarker levels, total and free, and patient age. The classification was based on 4 disease classes (D = 1) and a control class (D = 0) (Table 2).

Table 2. Biomarkers Data

| Diagnosis | Biomarker A | Biomarker B | Biomarker C |
|---|---|---|---|
| Disease A | 5,17 | 4,17 | 8,7 |
| Disease B | 0,83 | 0,91 | 0,88 |
| Disease C | 0,75 | 0,89 | 1,03 |
| Disease D | 0,81 | 0,86 | 1,12 |
| D = 0 | 0,89 | 0,95 | 0,96 |

**Search for informative signs.** To solve the problem of selecting the most informative features, it is advisable to consider the use of various statistical tests and measures of separability of classes. We assumed in our study that the null hypothesis says that the classes presented in Table 2 are inseparable, and the alternative that the classes are separable.

Statistical tests based on Student's t-Test and Mann-Whitney-Wilcoxon U-Test were aimed at pairwise comparison of two different classes. Therefore, a control class was selected and compared in pairs with disease classes. To evaluate the work of each statistical criterion, p-value is used - the probability of error when the null hypothesis is rejected [3].

For each test, 4 independent sets of comparisons of each type of diagnosis with the control class were formed. Several markers with a minimum p-value were selected from each sample. The resulting samples were combined into one set without repeating markers.

**Methodology and evaluation of classification efficiency.** Data classification was carried out in several stages: applying a sliding control scheme, randomly dividing the initial data sample into training and test data in equal proportions, training the classifier on the training sample and testing it on the test sample, assessing the quality of the classification using one of the described metrics.

A process was created containing classification methods: Logistic Regression, Support Vector Machine, Naive Bayes Classifier, K-Nearest Neighbors, Decision Tree, Random Forest.

In addition, a neural network model was taken separately for the study, trained on the same data and used to compare the experimental results. Model parameter settings (in Python) as shown in the following Table 3:

Table 3. Experimental Parameters

| Model | Parameter |
|---|---|
| Logistic Regression | max_iter=10, penalty=l2, solver=liblinear,tol=1e-4 |
| Support Vector Machine | decison_function_shape=ovo,C=1,kernel=rbf |
| Naïve Bayes | alpha=0.01 |
| K-Nearest Neighbors | n_neighbors=10 |
| Decision Tree | max_depth=3, min_samples_leaf=1, criterion=gini |
| Random Forest | n_estimators=10, max_depth=3, criterion=gini |
| Ensemble Learning | layer1=[LR,SVM,NB,KNN,DT,RF],layer2=[LR] |

**Application of the ensemble method.** The ensemble learning includes two stages: learn first-level, learn a second-level meta-classifier. We use the output of the first-level classifiers as the new features. Next, use the new features to train the second level meta-classifier (Figure 1).

As a result of the analysis, it can be argued that the k-nearest neighbors' method and the random forest method have the least efficiency in diagnosing. At the same time, the naive Bayesian method has the greatest efficiency, although its classification accuracy cannot be considered sufficient. Therefore, it was concluded that the most effective would be to use the ensemble method based on Bayes classifiers [4].

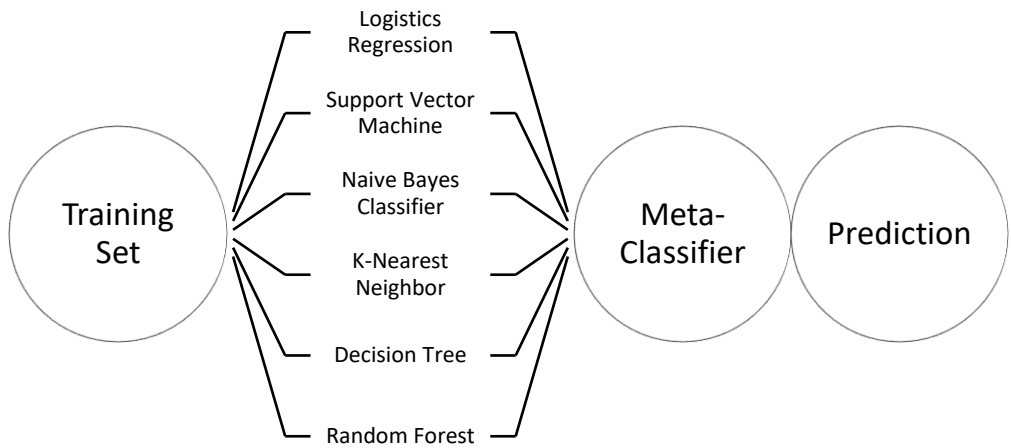

Figure 1. Two-levels ensemble approach

Our evaluation method consists of three parts: data processing, model training, and ensemble learning. In the data processing stage, we remove the stop words, remove punctuation, stemming [5].

In the model training stage, we select logistic regression as the meta-classifier to learn a second-level classifier. Before using ensemble learning, we need to set the hyperparameter of each classifier. We use train data and validation data to training each independent classifier, adjust the hyperparameters to achieve the best performance of the independent classifier on the validation set. The experiment results of each model on test data in the following Table 4:

Table 4. Experimental Parameters

| Model | Accuracy |
|---|---|
| Logistic Regression | 0.543 |
| Support Vector Machine | 0.432 |
| Naïve Bayes | 0.512 |
| K-Nearest Neighbors | 0.576 |
| Decision Tree | 0.534 |
| Random Forest | 0.489 |
| Ensemble Learning | 0.681 |

It can be seen from the table that ensemble learning has achieved the best performance and the performance of the decision tree in the performance of a single classifier is the best. Different classifiers can learn different data features, and ensemble learning can integrate the features learned by each classification and the advantages of each classifier. In addition, through experiments, we found that the performance of the logistic regression and support vector machines is stable, and the classification performance is not obviously different.

**Conclusions.** The proposed consolidating method for analyzing series of observations based on the fitted model allows one to study any number of markers. The ensemble included not only the listed classifier methods, but also the neural network, which showed comparable accuracy in the previous numerical experiment. The result of the concluded performance vector is very high, i.e., a fairly simple ensemble of methods is able to solve the problem more efficiently than individual methods.